\documentclass[showpacs,a4paper,twocolumn,superscriptaddress,floatfix]{revtex4-1}

\usepackage{times}
\usepackage{graphicx}
\begin{document}

\title{Hot entanglement in a simple dynamical model}

\bigskip

\author{S.\ Scheel, J.\ Eisert, P.\ L.\ Knight, and M.\ B.\ Plenio}

\affiliation{QOLS,
Blackett Laboratory, Imperial College of Science, Technology 
and 
Medicine, London, SW7 2BZ, UK}

\date{\today}

\begin{abstract}
How mixed can one component of a bi-partite system be initially
and still become entangled through interaction with a thermalized partner?
We address this question here. In particular,
we consider the question of how mixed a two-level system and a
field mode may be such that free entanglement arises in the course
of the time evolution according to a Jaynes-Cummings type
interaction. We investigate the situation for which the
two-level system is initially in mixed state taken from a one-parameter
set, whereas the field has been prepared in an arbitrary 
thermal state. Depending on the particular choice for the
initial state and the initial temperature of the quantised
field mode, three cases can be distinguished: (i)  free entanglement 
will be created immediately, (ii) free entanglement will be generated, but only at a later
time different from zero, (iii) the partial
transpose of the joint state remains positive at all times.
It will be demonstrated that   increasing
the initial temperature
of the field mode may cause the joint state to 
become distillable
during the time evolution, in contrast to a non-distillable
state at lower initial temperatures.
We further assess the generated entanglement quantitatively, by
evaluating the logarithmic negativity numerically, 
and by providing an analytical upper bound.

\end{abstract}

\maketitle

\section{Introduction}

Interacting quantum systems typically develop correlations
in the course of their dynamics, either of classical or
of genuinely quantum nature \cite{Dynamical,Purity,Jens,Gemmer}. 
If two systems are initially in a 
pure product state which does not decouple from the interaction 
Hamiltonian, entanglement will always occur. Depending on the
particular type of the interaction and the choice of the initial
state, it may well be that the degree of entanglement being generated
is very small indeed for a rather long time period \cite{Gemmer}. 
Entanglement is nevertheless
an unavoidable byproduct of the time evolution.
If the joint system is initially in a product
state, but with one or both subsystems being mixed, then it
is not necessarily clear that entanglement will arise at any
point in the dynamics. For finite-dimensional systems it
is obvious  that whenever the joint system is suitably mixed initially,
then the joint state will stay separable at all times. For infinite-dimensional
systems, however, a more complex behaviour may be anticipated. Intuitively,
one may still expect that if both systems are initially 
very mixed, then the bi-partite 
state should not develop entanglement
at any time. In the case that 
one system is in an arbitrarily mixed state
and the other is pure, then it is conceivable that entanglement
is always
generated. This has in fact been demonstrated in the case 
of a Jaynes-Cummings-type interaction \cite{Jaynes,Jaynes2,VogelWelsch}
of a two-level system
coupled to a field mode \cite{Purity}, and in the Caldeira-Leggett
model for quantum Brownian motion, where a single field mode
is considered that is linearly coupled to many other field modes
initially prepared in some thermal state \cite{Jens}.

The Jaynes-Cummings model is also particularly suitable to assess
the situation when initially both systems are initially in a mixed state.
This model describes in the most basic way the interaction
between light and matter: it models the dynamics under
linear coupling of a two-dimensional system to a near resonant
quantised 
single field mode with
a rotating-wave Hamiltonian. The Jaynes-Cummings model can be analytically
solved in the sense that the time evolution can be explicitly
evaluated for all times. Hence, one may investigate 
the trade-off between the
degree of mixing of the two-level system and the mixing of the field 
mode. 

\begin{figure}
\includegraphics[width=.42\textwidth]{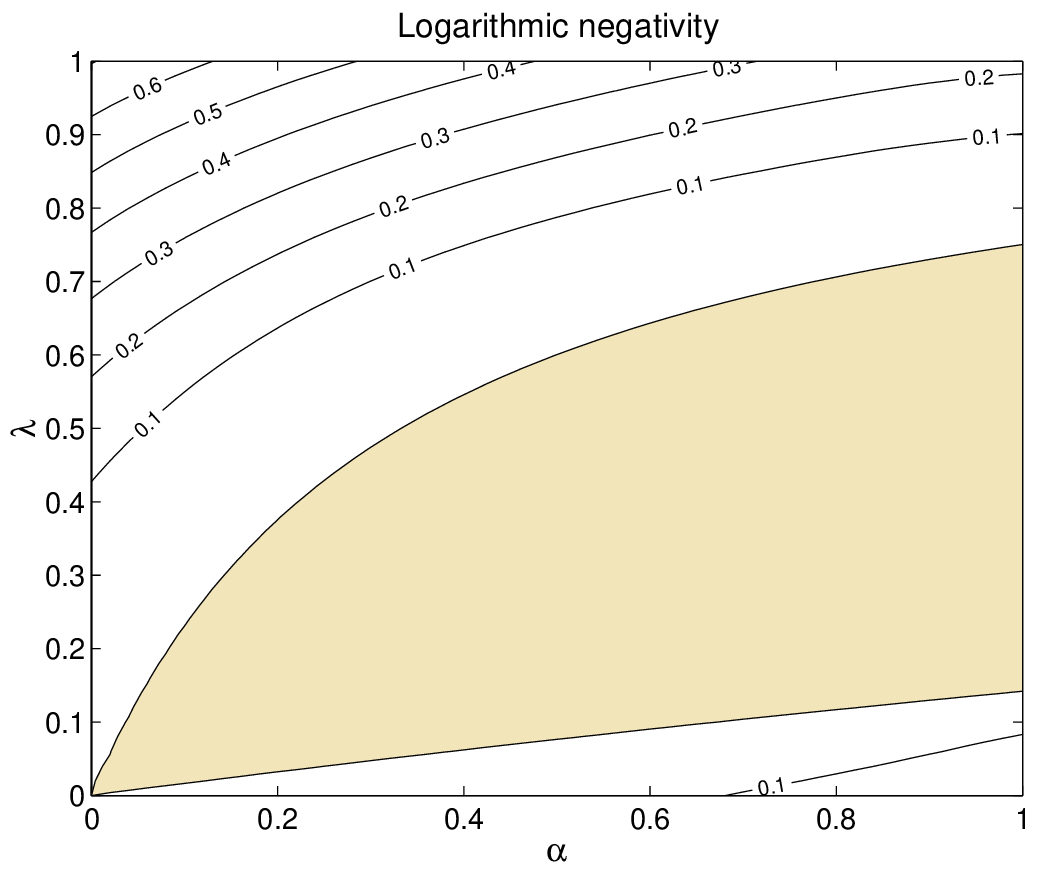}
 \caption{This figure depicts the maximal degree of entanglement
that arises in the joint state of a two-level system coupled to a single
field mode in the Jaynes-Cummings model. The field mode is initially
in a thermal state, the two-level
system is prepared in $\lambda|e\rangle\langle e|+ (1-\lambda) |g\rangle\langle g|$,
$\lambda\in[0,1]$. Zero temperature corresponds to $\alpha=0$, and the limit
of infinite temperature is reflected by the limit $\alpha\rightarrow 1$.
The degree of entanglement is measured in terms of the logarithmic negativity.
The shaded area represents the situation where the log-negativity is zero
for all times, meaning that the joint state has a positive partial transpose
at all times. In this figure, 
$\alpha$ is related to the initial
inverse temperature $\beta$ of the
field mode according to $\alpha= m/(m+1)$, where $m= (\exp(\beta)-1)^{-1}$
is the mean photon number.}
\end{figure}

In our analysis, we consider the case where the field mode
is initially in a thermal state. The set of initial states of the two-level system, 
from now on labeled as $A$, will be given by
\begin{equation}
	\rho_A=\lambda |e\rangle\langle e |
	+ (1-\lambda) |g\rangle\langle g|, \,\,\,\lambda\in[0,1],
\end{equation}
where $|e\rangle$ and $|g\rangle$ denote the state vectors corresponding to the
excited and the ground state, respectively. We will analyse the entanglement
properties of the joint system at all times, for all initial temperatures
of the field mode and all $\lambda\in[0,1]$.
 It will turn out that 
depending on
the actual parameters in the model,  three regimes can be distinguished:
the regime where (free) entanglement is generated immediately, the regime where
the joint state will be entangled at some later time $t>0$, and the
regime where the joint state has a positive partial transpose for all times.
The regime where entanglement will immediately be created is by no means
restricted to the case where the two-level system is initially pure.
At arbitrarily large temperature of the field mode one may still have 
a very mixed initial state of the two-level system -- yet different from the
maximally mixed state -- in order to obtain an entangled state for all times.

\section{Jaynes-Cummings Model for Atom-Field Interaction}

The Jaynes Cummings interaction \cite{Jaynes,Jaynes2} of the near
resonant interaction with a single mode quantised field with
annihilation and creation operators $a$ and $a^\dagger$, respectively,
can be written as
\begin{equation}
	H_{int}= \gamma \left( |e\rangle\langle g| a + |g\rangle\langle e| a^\dagger
	\right).
\end{equation}
For simplicity, we set $\hbar=1$ and assume the
frequency of the field mode to be unity as well, which does
not imply a restriction of generality for our argument.
The field mode, associated with the label $F$, is initially prepared
in a thermal state with respect to the inverse temperature $\beta>0$, i.e,
\begin{equation}
	\rho_F= \frac{\exp(-\beta H_F)}{\text{tr}[\exp(-\beta H_F)]}
	= \sum_{n=0}^\infty p_n |n\rangle\langle n|,
\end{equation}
where $H_F:= a^\dagger a+1/2$ is the Hamiltonian of $F$
and
\begin{equation}
	p_n:= \frac{m^n}{(1+m)^{n+1} }.
\end{equation}
$m$ is the mean photon number at the inverse temperature
$\beta$, and is given by
\begin{equation}
	m:= \left( \exp(\beta)-1\right)^{-1}.
\end{equation}
The joint initial state $\rho_S\otimes \rho_F$ evolves
in the interaction picture
according to
\begin{equation}
	\rho(t)= U(t) ( \rho_S\otimes \rho_F)  U(t)^\dagger, \,\,t\geq 0,
\end{equation}
where $U(t):= \exp(-i H_{int} t)$. The formal time evolution can be
made explicit, as the Jaynes-Cummings model
can be analytically solved.
The state $\rho(t)$ 
at a time $t\geq 0$ can be evaluated
using the standard techniques to solve the Jaynes Cummings model
\cite{VogelWelsch}.
For brevity, we will write $\rho$ instead of $\rho(t)$ from now on.
With the above notation, one arrives at
\begin{equation}
	\rho = \sum_{n=0}^\infty p_n \rho_n,
\end{equation}
where
\begin{eqnarray}
	\rho_n  = &\lambda & \Bigl[
	c_n^2 |e,n\rangle\langle e,n|
	+ s_n^2 |g,n+1\rangle\langle g,n+1| \\
	&+& i c_n s_n \left( |g,n+1\rangle\langle e,n|
	-
	| e,n\rangle\langle g,n+1 |\right)\nonumber\\
	+ (1 & - &\lambda) \Bigl[
		s_{n-1}^2 |e,n-1\rangle\langle e,n-1|+
		c_{n-1}^2 
		|g,n\rangle\langle g,n|
		\nonumber\\
	&+& i c_{n-1} s_{n-1}\left(
		| e,n-1\rangle\langle g,n |- 
		|g,n\rangle\langle e,n-1| \right) 
	\Bigr],\nonumber	
\end{eqnarray}	
if $n>0$, and
\begin{eqnarray}
	\rho_0 & = & \lambda  \Bigl[
	c_0^2 |e,0\rangle\langle e,0|
	+  s_0^2|g,1\rangle\langle g,1|\\
	&-& i c_0 s_0 | e,0\rangle\langle g,1 |
	+ i c_0 s_0 |g,1\rangle\langle e,0|\Bigr]\nonumber\\
	&+& (1 -\lambda)  |g,0\rangle\langle g,0| .\nonumber	
\end{eqnarray}	
The symbol 
\begin{equation}
	\omega_n:= 2 \gamma (n+1)^{1/2}
\end{equation}
has been used for the Rabi frequency, and $s_n$ and $c_n$ are defined
as
	$c_n:= \cos(\omega_n t/2)$, and $s_n:= \sin(\omega_n t/2)$
for $n=0,1,2,...$.
We will set $\gamma=1$,
again for reasons of clarity,
which simply corresponds to a rescaling of the units of time.

\section{Entanglement Properties and Partial Transpose}

In order to assess the entanglement properties
of the joint state $\rho$ at a later time $t>0$ we
investigate the spectrum of the partial transpose  $\rho^\Gamma$ of 
$\rho$. This partial transpose takes the form
of a direct sum, such that the spectrum of $\rho^\Gamma$ can be
evaluated from the spectrum of each block. In order to simplify
the notation, we set 
\begin{equation}
	p_n=(1-\alpha) \alpha^n,
\end{equation}
with $\alpha\in [0,1)$. 
The partial transpose is then explicitly given by
\begin{eqnarray}
	\rho^\Gamma 
	&=& (1-\alpha) \sum_{n=0}^\infty\alpha^{n-1 }  
	\Bigl[
	(\alpha c_n^2 \lambda  + \alpha^2
	s_n^2 (1-\lambda) )  |e,n\rangle\langle e,n|\nonumber\\
	&+& 	
	i (\lambda\alpha  -   (1-\lambda)\alpha^2 )	
	\nonumber \\
	&\times &
	c_{n-1} s_{n-1} (
	|e,n+1\rangle\langle g,n|
	-
	|g,n\rangle\langle e,n+1 |)\nonumber\\
	&+& ( s_{n-1}^2 \lambda + (1-\lambda) c_{n-1}^2 \alpha)
	|g,n\rangle\langle g,n| 
	\Bigr].
\end{eqnarray}
From this equation, the spectrum $\sigma(\rho^\Gamma)$ of $\rho^\Gamma$
can be read off: it is given by
\begin{equation}
	\sigma(\rho^\Gamma) = 
	\left\{ 1-\alpha, 
	\lambda_0^+, \lambda^-_0 , \lambda_1^+, \lambda^-_1, ... \right\},
\end{equation}
where $\lambda_n^+,\lambda^-_n$, $n=0,1,2, ...$, with $\lambda_n^+\geq \lambda_n^-$
are the two eigenvalues of the $2\times 2$-matrices
\begin{equation}\label{Mn}
	M_n:=\left(
	\begin{array}{cc}
	A_n& C_n \\
	C_n^\ast & B_n\\
	\end{array}
	\right),
\end{equation}
with entries
\begin{eqnarray}
	A_n&:=&  (1-\alpha)\alpha^{n+1}
	(c_{n+1}^2 \lambda + \alpha s_{n+1}^2 (1-\lambda) ), \\
	B_n&:=& (1-\alpha) \alpha^{n-1}  
	(\alpha c_{n-1}^2 (1- \lambda) + s_{n-1}^2 \lambda),\\
	C_n & :=& i (1-\alpha) \alpha^{n}( \lambda - \alpha(1-\lambda) )
	c_{n} s_{n}
\end{eqnarray}
if $n>0$, and
\begin{eqnarray}
	A_0&:=&  (1-\alpha)\alpha
	(c_{1}^2 \lambda + \alpha s_{1}^2 (1-\lambda) ), \\
	B_0&:=& (1-\alpha)  (1- \lambda),\\
	C_0 & :=& i (1-\alpha) ( \lambda - \alpha(1-\lambda) )
	c_{0} s_{0}.
\end{eqnarray}
The remaining task is to identify those values for $\alpha$ and
$\lambda$ for which there exists a $t>0$ such that at least one
$\lambda_n^-$ is negative. Due to the form of $M_n$
it is obvious that $\lambda_n^+>0$ for all $n=0,1,2,...$. Hence, it 
is sufficient to consider the positivity of $\det[M_n]$. We first consider the
case that $n>0$. 
The numbers 
\begin{equation}
n^{1/2},\, (n+1)^{1/2},\, (n-1)^{1/2}
\end{equation} 
are not in a rational relationship to each other. Therefore,
each contribution to $\text{det}[M_n]$ may be individually
minimised. Hence, there always exists a time
$t>0$ such that $M_n$ is given by Eq.\ (\ref{Mn}) with
\begin{eqnarray}
	A_n&=&  (1-\alpha)\alpha^{n+1} f(\alpha,\lambda),\\
	B_n&=& (1-\alpha) \alpha^{n-1}  f(\alpha,\lambda),\\
	C_n & =& i (1-\alpha) \alpha^{n}( \lambda - \alpha(1-\lambda) )/2,
\end{eqnarray}
where
\begin{equation}
	f(\alpha,\lambda):=
	\min( \lambda, \alpha  (1-\lambda)).
\end{equation}
This choice of $t$ corresponds to the 
sitation where $\text{det}[M_n]$ takes its
minimal value on the 
interval $t\in(0,\infty)$. 
This means that there exists a time $t>0$ with
$\det[M_n]<0$ if and only if
\begin{eqnarray}\label{final}
	f(\alpha,\lambda)^2 - ( \lambda - \alpha(1-\lambda) )^2/4<0.
\end{eqnarray}
In the case $n=0$, then one arrives with an analogous argument at the statement
that there exists a time $t>0$ such that $\det[M_0]<0$ if and only if
\begin{equation}\label{final2}
	\alpha 
	f(\alpha,\lambda)(1-\lambda) - ( \lambda - \alpha(1-\lambda) )^2/4<0.
\end{equation}
Eqs.\ (\ref{final}) and (\ref{final2})
provide the desired relationship between $\alpha$ and $\lambda$.
Three cases may now be distinguished:
\begin{itemize}
\item[(i)] The values of $\lambda\in[0,1]$ and $\alpha\in[0,1)$
are such that
neither Eq.\ (\ref{final}) nor  Eq.\ (\ref{final2}) are
satisfied. This means that the joint state of the two-level system
and the field mode has a positive partial transpose {\it for all times}
(i.e., it is a PPT state).
\item[(ii)] The values of $\lambda$ and $\alpha$ 
are such that Eq.\ (\ref{final2}) is satisfied,
but Eq.\ (\ref{final}) is not satisfied. This means that {\it there exists
a time $t>0$} such that the joint state $\rho$
has a non-positive partial transpose
(i.e., it is a NPPT state). This time can easily be evaluated: If 
\begin{equation}
\lambda\leq \alpha (1-\lambda), 
\end{equation}
then 
it is given by the time
$t>0$ that achieves 
$c_0 s_0=1/2$ and $c_1=1$. If $\lambda > \alpha (1-\lambda)$,
then $c_0 s_0=1/2$ and $s_1=1$ have to be simultaneously
satisfied.

\item[(iii)] The values of $\lambda$ and $\alpha$
are such that Eq.\ (\ref{final}) is
fulfilled. Then, again, the joint state will become a NPPT state.
However, now the situation is different compared to case (ii): 
for any time $t>0$ one can find an $n=1,2,...$ such that 
$\text{det}[M_n]<0$. This means that the joint state $\rho$
becomes an NPPT state {\it instantaneously}, or, more precisely,
$\rho$ is an NPPT state for all times $t\in (0,\infty)$.

\end{itemize}

It is important to note that 
whenever the state is a NPPT state, it is also a free entangled one.
This can be seen when considering 
\begin{equation}
	(1\otimes \pi_n)\rho (1\otimes \pi_n)
\end{equation}
for $n=1,2,...$, where
\begin{equation}
	\pi_n:=\sum_{i=n-1}^{n+1} |i\rangle\langle i|.
\end{equation}
This is for each $n$
a projection onto a $2\times 3$-dimensional 
subspace. 
By investigating the block diagonal 
structure of $\rho^\Gamma$, 
one finds that the partial transpose of 
$(1\otimes \pi_n)\rho (1\otimes \pi_n)$
has again a direct sum structure. 
It follows from direct investigation of 
this projection that
 $\rho^\Gamma$
is non-positive if and only if there exists an $n$ such that
$(1\otimes \pi_n)\rho (1\otimes \pi_n)$ is non-positive. Hence,
whenever one concludes that $\rho$ is NPPT, then it is also free entangled,
as for $2\times 3$-systems being NPPT and being distillable 
are equivalent \cite{bound}.

The situation that has just been described
is depicted in Figs.\ 1 and 2. In Fig.\ 1,
the white area corresponds to the situation
where for some time $t>0$
a free entangled 
state is reached, which is case (ii). 
The shaded area is assigned case (i), the
PPT case. The lines in Fig.\ 2 correspond to 
equality in Eq.\ (\ref{final}): then free entanglement is created immediately,
as in case (iii).

It is interesting to note that even in the limit $\alpha\rightarrow 1$,
which physically is the limit of infinitely large temperature,
encounters a PPT
state. No matter how large the temperature, there always exists a value
$\lambda\in(0,1)$
such that the joint state becomes entangled.
The intersection of the parametrised
curve of equality in Eq.\ (\ref{final}) with
$\alpha=1$ associated with 'infinite temperature' can be easily evaluated. 
One finds two intersection points at $\lambda=3/4$ and $\lambda=1/4$. 
The intersection points of the curve of equality in Eq.\ (\ref{final2}) with
$\alpha=1$ are given by $\lambda = 1/2 - \sqrt{2}/4= 0.14645$ and
$\lambda=3/4$. That is, whenever $\lambda$ is larger than $3/4$
or smaller than $ 1/2 - \sqrt{2}/4$,
then 
one can be sure that distillable entanglement will be generated. 
To this extent, both the two-level system and the field mode 
may therefore be mixed initially. If $\lambda=1/2$, which corresponds
to the maximally mixed state for the two-level system, there exists a 
temperature for which there is no distillable entanglement left.
Another interesting situation arises when $\lambda=1/10$ is chosen:
Then, at small temperatures, distillable entanglement will be generated.
As the temperature increases, that is, 
with increasing $\alpha$,
no free entanglement will occur. But then again, at very large temperatures
free entanglement will again arise. 

\begin{figure}
\includegraphics[width=.42\textwidth]{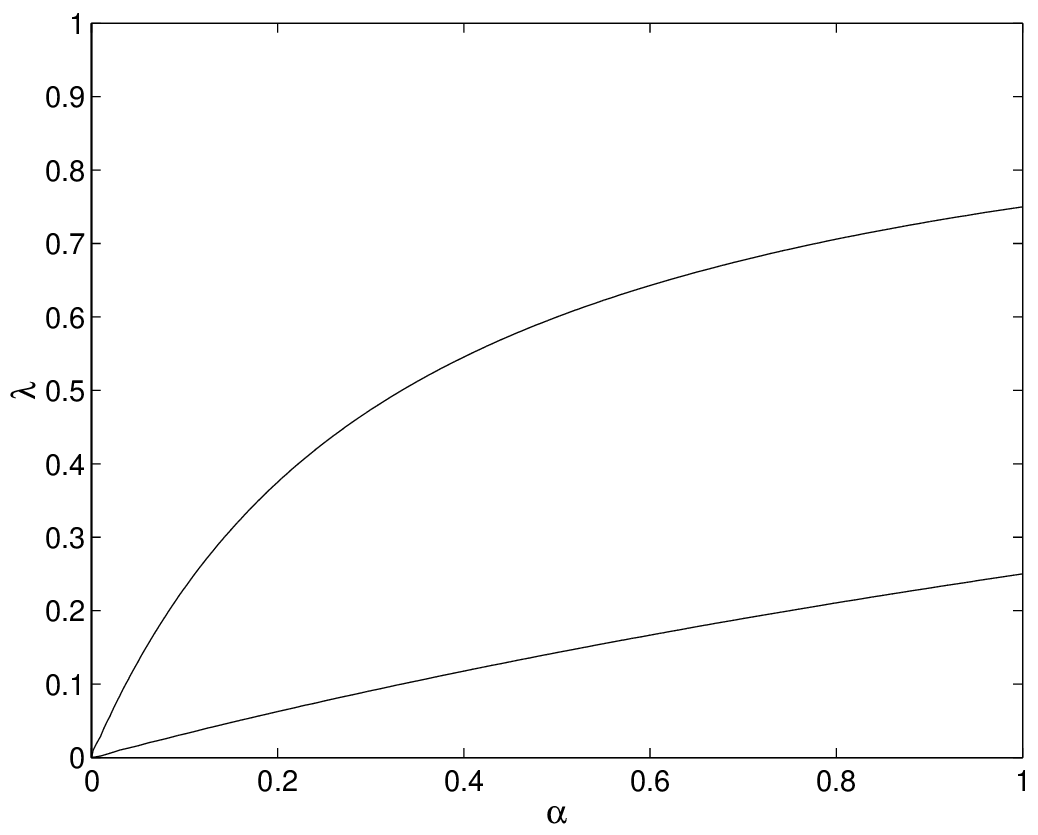}
 \caption{
This figure depicts the boundary between the region
where free entanglement is created immediately and the region for which 
there exists a $t>0$ such that the joint state is PPT at time $t$.}
\end{figure}

From the above considerations, one can also find an upper bound for the 
distillable entanglement with respect to local operations with
classical communication 
that can maximally
be achieved in the course of the time evolution.
The log-negativity is such an upper bound for the distillable
entanglement. The negativity \cite{neg,vidal}
of a state $\rho$ is defined
as $\| \rho^\Gamma \|_1-1$, where $\|.\|_1$ denotes 
the trace-norm; it 
has been
shown to be an entanglement monotone \cite{vidal,phd}. 
The log-negativity is
nothing but $N(\rho):= \log_2 \| \rho^\Gamma \|_1$; it is an easily computable 
upper bound for the distillable entanglement \cite{vidal}. 
Fig.\ 1 depicts the maximal log-negativity with respect to all times
from numerical investigations. Note that around 
$\lambda=1/10$, the
counterintuitive situation occurs that 
the maximal degree of entanglement that is 
achieved {\it increases}\/ 
with the initial temperature of the field mode. 

An upper bound for the infimum
$N(\rho):= \log \| \rho^\Gamma \|_1$ over all times can be evaluated from the
above explicit expression for $\rho^\Gamma$. First, note that
\begin{equation}
	\|\rho^\Gamma\|_1 = 1+ 2\sum_{n=0}^\infty |\lambda_n^-|. 
\end{equation}
An upper bound for the log-negativity and therefore for the
distillable entanglement can then be evaluated from
the extremal sitations in each block of $\rho^\Gamma$. One arrives
at
\begin{eqnarray}
	N(\rho)\leq \log_2 ( 1+2 \min(0, |\mu^- |) + 2\min (0, | \nu^-|)  ),
\end{eqnarray}
where $\mu^- $ and $\nu^-$ are 
the smallest eigenvalues of 
\begin{eqnarray}
	(1-\alpha)
	\left(
	\begin{array}{cc}
	\alpha f(\alpha,\lambda) &  (\lambda - \alpha(1-\lambda))/2\\
	 (\lambda - \alpha(1-\lambda))/2 & 1-\lambda
	\end{array}
	\right)
\end{eqnarray}
and
\begin{eqnarray}	
	\left(
	\begin{array}{cc}
	\alpha^2 f(\alpha,\lambda) & \alpha (\lambda - \alpha(1-\lambda))/2\\
	\alpha (\lambda - \alpha(1-\lambda))/2 & f(\alpha,\lambda)
	\end{array}
	\right),
\end{eqnarray}
respectively.
A particularly simple situation is the case that $\alpha=1$, corresponding
to 'infinite temperature' of the field mode. 
Then one finds that
\begin{equation}
	N(\rho)\leq \left\{
	\begin{array}{ll}
	\log_2(2-4\lambda),& \text{for }\lambda\in[0,1/4],\\
	\log_2(4\lambda-2),& \text{for }\lambda\in[3/4,1],\\
	0 & \text{otherwise.}
	\end{array}
	\right.
\end{equation}

\section{Conclusions and Outlook}

In the main part of this paper we have investigated the
properties of the dynamically emerging entanglement
in the one-photon Jaynes-Cummings model. 
We have found that three different regimes occur, depending
on the actual initial joint product state: there may be free 
entanglement at all times, and only at a later time. Also, there is
the case that the partial transpose of the joint state stays positive
at all times, such that the state does not contain
distillable entanglement 
\cite{bound}. 
That is, the state is then either separable or
bound entangled.
One should, however, keep in mind that for bi-partite
systems, where at least one
part is infinite-dimensional
(such as the system considered in this paper), some care is
needed when speaking about the separability of a state:
the set of 
non-separable states is
then 
trace-norm dense in the set of all states 
\cite{Halvorson} (see also Refs.\ \cite{Pawel}): 
this means that for any separable state
there exists an entangled
state arbitrarily close to the original state as measured by 
the trace-norm, a state that leads consequently 
to arbitrarily similar probability distributions in quantum 
hypothesis testing distinguishing the two states. 
In contrast, there are clearly entangled neighbourhoods
of entangled states. 

One -- possibly not very surprising -- principal observation is that 
the point corresponding to $\lambda=1/2$ and $\alpha=1$ 
in Fig.\ 1 is contained in the shaded region. This means that 
there always exists a temperature such that if {\it both}
the two-level system $S$ and the field mode $F$
are initially in thermal states corresponding to this
temperature, then the composite system 
will certainly not develop free
entanglement. The specific shape of the shaded area 
in Fig.\ 1 is clearly a property of the considered 
one-photon Jaynes-Cummings model. By appropriately choosing
the interaction Hamiltonian -- for example by introducing a $k$-photon
Jaynes-Cummings interaction -- 
one may obviously be in the  
position to change the shape of this area, 
and even to make its area 
very small. If one allows for a time
dependence of the interaction, one can also find Hamiltonians
such that for {\it all} 
initial temperatures of both parts of a bi-partite
system, there will be a time $t>0$ such that the joint system
will eventually
become free entangled. 
This leads to the general question whether this connection is
unavoidable: there may even be bi-partite
infinite-dimensional systems
with the property that (i) the thermal states exist for
all temperatures for both parts of the system and (ii) with
the property that the composite system becomes entangled
for all initial temperatures for a fixed time-independent 
interaction.  The existence of such systems may not appear
particularly plausible. A complete answer to this question may
however lead to a more thorough understanding of the exact link
between the initial mixedness of the state 
of a composite quantum system and the capability 
of producing entanglement in the course of the time evolution.

\section{Acknowledgements}

This work has been supported by the 
European Union Networks
EQUIP, QUEST, and QUBITS, 
the EPSRC, the DFG 
 (Schwer\-punkt\-programm QIV), 
and the A.-v.-Humboldt-Foundation
(Feodor-Lynen-Grants of JE and SS).

\end{document}